\documentclass[conference]{IEEEtran}
\IEEEoverridecommandlockouts
\usepackage{cite}
\usepackage{amsmath,amssymb,amsfonts}
\usepackage{graphicx}
\usepackage{textcomp}
\usepackage{xcolor}
\usepackage{hyperref}
\usepackage{bm}
\usepackage{subcaption}
\usepackage{textcomp}%
\usepackage{manyfoot}%
\usepackage{booktabs}%
\usepackage{algorithm}%
\usepackage{algorithmicx}
\usepackage{algpseudocode}%
\usepackage{listings}%
\usepackage{nicematrix}
\usepackage{lineno}
\usepackage{tikz}
\usetikzlibrary{decorations.pathreplacing,calc}

\usepackage[section]{placeins}
\def\BibTeX{{\rm B\kern-.05em{\sc i\kern-.025em b}\kern-.08em
    T\kern-.1667em\lower.7ex\hbox{E}\kern-.125emX}}
\begin{document}

\title{Role-Selection Game in Block Production under Proposer-Builder Separation\\
}

\author{\IEEEauthorblockN{Yanzhen Li}
\IEEEauthorblockA{\textit{School of Systems Science} \\
 \textit{Beijing Normal University}\\
 Beijing, China \\
 liyanzhen@mail.bnu.edu.cn}
 \and
 \IEEEauthorblockN{Zining Wang}
 \IEEEauthorblockA{\textit{School of Engineering Mathematics and Technology} \\
 \textit{University of Bristol}\\
 Bristol, United Kingdom \\
 zining.wang@bristol.ac.uk}
 }

\maketitle

\begin{abstract}
To address the risks of validator centralization, Proposer-Builder Separation (PBS) was introduced in Ethereum to divide the roles of block building and block proposing, fostering a more equitable and decentralized block production environment. PBS creates a two-sided market in which searchers submit valuable bundles to builders for inclusion in blocks, while builders compete in auctions for block proposals. In this paper, we formulate and analyze a role-selection game that models how profit-seeking participants in PBS strategically choose between acting as searchers or builders, using a co-evolutionary framework to capture the complex interactions and payoff dynamics in this market. Through agent-based simulations, we demonstrate that agents’ optimal role—acting as searcher or builder—responds dynamically to the probability of conflict between bundles. Our empirical game-theoretic analysis quantifies the equilibrium frequencies of role selection under different market conditions, revealing that low conflict probabilities lead to equilibria dominated by searchers, while higher probabilities shift equilibrium toward builders. Additionally, bundle conflicts have non-monotonic effects on agent payoffs and strategy evolution. Our results advance the understanding of decentralized block building and provide guidance for designing fairer and more robust block production mechanisms in blockchain systems.

\end{abstract}


\section{Introdution}
Decentralized Finance (DeFi), which employs blockchain-based smart contracts to deliver financial services similar to traditional systems \cite{buterin2014next}, reveals a weakness in its inability to maintain the sequence of transactions, affecting the outcomes of contract execution. 
As an example, attackers in decentralized exchanges (DEXs) can exploit the public mempool by spotting valuable transactions and placing higher-fee orders to engage in front-running, ensuring their transaction gets executed first. Worse yet, they can employ sandwich attacks, combining front-running and back-running to manipulate victim transactions for considerable financial gain \cite{zhou2021}. 
Such potential revenue for block producers that can be extracted by strategic transaction selection or reordering is termed as \textit{miner/maximal extractable value} (MEV) \cite{daian2020flash}.

It is worth noting that MEV is not solely about financial gains and losses; it also has posed a threat to decentralization. The fundamental principle of blockchain technology is that no small group of entities should be able to manipulate the blockchain's records or impose censorship \cite{grandjean2023ethereum}.  In the current Ethereum Proof of Stake (PoS) paradigm, traditional miners in the Proof of Work (PoW) are replaced by validators, and these block producers can allocate MEV rewards as additional stakes to block reward, bolstering their power over the protocol. However, the asymmetry in block producers' capacity to extract MEV constitutes a long-term force driving centralization \cite{bahrani2024centralization,capponi2024proposer}. 
As a countermeasure to this trend, the Ethereum community introduced \textit{Proposer-Builder Separation} (PBS) \cite{vitalikPBS} to ensure a fairer landscape.
The logic behind PBS is to split the role of the validators into two parts, separating block building from block proposal. Specialized entities, known as block builders, are responsible for creating the most profitable blocks, while the block proposer (the chosen PoS validator) selects the block with the highest bid through an auction\footnote{Flashbots' MEV-Boost building auction exemplifies a successful out-of-protocol PBS implementation, with approximately 90\% of all new blocks being created via the MEV-Boost block \cite{oz2024wins}.} among builders to propose it on the blockchain. This not only creates opportunities to prevent transaction censorship at the protocol level but also levels the MEV extraction ability between hobbyists and institutional validators.

Despite the introduction of PBS to tackle the issue, it appears ineffective in resolving the problem. Currently, three builders—beaverbuild, rsync, and Titan (BRT)—are responsible for constructing nearly all the blocks \cite{oz2024wins,yang2024decentralization}, which continues to compromise Ethereum’s censorship resistance and decentralization.
Moreover, such concentration creates a kind of barrier to entry for new builders, exemplified by a ``chicken-and-egg'' problem \cite{oz2024wins,Titan,yang2024decentralization}.
Decentralized block building has become the primary improvement target of the current PBS framework, wherein searchers and builders can collaborate to construct the optimal block together by leveraging cryptographic technologies, such as Trusted Execution Environments \cite{BlockbuildingTEE} and Multi-Party Computation \cite{shi2023can}.
The concept of such a decentralized block building naturally prompts inquiries regarding how each profit-seeking actor can derive benefits from the block-building process. To address this question, it is essential to model the strategies employed by these profit-seeking actors.

In this work, we study a strategic interaction wherein agents endowed with profit opportunities can choose between two roles: acting as searchers, who share a portion of their profit opportunities with builders to collaboratively construct blocks and thereby avoid involvement in the block building auction, or as builders, who participate in the auction under the PBS framework in pursuit of surplus. We model this block building game as a complex adaptive system~\cite{holland1992adaptation}, where agents continuously optimize their strategies using a genetic algorithm inspired by reinforcement learning principles. Through agent-based simulation, we investigate the co-evolutionary dynamics of agents' strategies, demonstrating the feasibility and rich behavioral implications of our proposed framework.\footnote{The code supporting the findings of this paper is available at \href{https://github.com/EthanLIYanzhen/Role-Selection-Game-in-Block-Production-under-Proposer-Builder-Separation.git}{https://github.com/EthanLIYanzhen/Role-Selection-Game-in-Block-Production-under-Proposer-Builder-Separation.git}}

The key contributions of this study are summarized as follows:

\begin{enumerate}
    \item We consider a two-sided market in our model, inherently more challenging to manage than a one-sided market, involving bundle interactions as evidenced in the block building process utilizing a greedy algorithm. In this model, agents' strategies are not fixed arbitrarily; they emerge from a learning process based on a co-evolution supported by genetic algorithms. 
    \item Our agent-based simulations enable us to clearly observe the co-evolutionary outcomes of the agents. In addition, we analyze the impact of bundle interactions on the evolution of agents' strategies and the distribution of benefits among participants. Our findings reveal a non-monotonic effect of the conflict probability between bundles on the co-evolutionary outcomes of the system.
    \item We consider the role selection of agents in the block building process as a meta-game, which encompasses meta-strategies of block building and bundle sharing. Based on this, we conduct an empirical game-theoretic analysis (EGTA) \cite{tuyls2018generalised} and calculate the dynamic equilibrium solutions of agents' strategies using the $\alpha$-Rank \cite{omidshafiei2019alpha} method, enabling us to predict the landscape of the block building process across different market conditions. 
\end{enumerate}







\section{Related Works}\label{sec:relatedwroks}
\subsection{Transaction Fee Mechanisms}
Certain researchers employ mechanism design theory to evaluate the feasibility of establishing a dominant-strategy incentive-compatible transaction fee mechanism \cite{roughgarden2020transaction,roughgarden2021transaction,bahrani2024transaction,bahrani2023transaction,chung2023foundations}. However, the impossibility results derived from these theoretical studies indicate that the desired attributes of an ideal transaction fee mechanism cannot be achieved simultaneously.
For instance, Bahrani et al. \cite{bahrani2023transaction} illustrate that no fee mechanism can attain incentive compatibility if block producers strive to actively extract MEV, as commonly observed in practice.
The analysis in \cite{bahrani2024transaction}, focused on a transaction fee mechanism (TFM) problem for a model featuring searchers and block producers (builder-proposer integration in PBS), considers a more detailed block production process, particularly a decentralized version of PBS.
Unlike the literature grounded in mechanism design, this paper primarily focuses on the strategies of participants within the MEV supply chain under the more practical framework of non-revealing mechanisms, specifically the first-price auction.

\subsection{Empirical Studies on PBS}

In recent literature, several studies have investigated the PBS landscape, underscoring the mounting tendency towards centralization of block building within the framework of PBS.
Heimbach et al. \cite{heimbach2023ethereum} provide a comprehensive empirical analysis of Ethereum’s PBS adoption, revealing persistent centralization among builders and relays, while also highlighting that PBS, despite intentions, has not effectively mitigated censorship and often relies on the trustworthiness of relays.
Gupta et al. \cite{gupta2023centralizing} empirically demonstrate that builders proficient in capturing CEX/DEX arbitrage (e.g., HFT firms) can construct blocks with significant top-of-block value. 
Yang et al. \cite{yang2024decentralization} propose two metrics to quantify the competitiveness and efficiency of the MEV-boost auction, highlighting issues related to current PBS, including entry barriers for builders and trust crises faced by searchers.
{\"O}z et al. \cite{oz2024wins} reveal that builders' profitability is correlated with access to exclusive order flow, which is subsequently linked to their market share,  thereby naturally underscoring a ``chicken-and-egg" dilemma.

\subsection{Game-Theoretic Approaches to PBS}
Some scholars have performed theoretical analyses of the PBS framework using game-theoretic models. 
Bahrani et al. \cite{bahrani2024centralization} assess the effectiveness of PBS, quantifying how the competitiveness of the builder ecosystem influences the ability of PBS to mitigate reward heterogeneity among proposers.
In addition to their empirical work discussed earlier, Gupta et al. \cite{gupta2023centralizing} also develop a game-theoretic model illustrating how builders with superior top-of-block capabilities can reinforce their dominance in order flow and PBS auctions through a positive feedback mechanism.
Capponi et al. \cite{capponi2024proposer} propose a three-stage game-theoretical model involving heterogeneous builders and a single order flow provider, further elucidating the mechanisms underlying builder dominance.
Additionally, Mamageishvili et al. \cite{mamageishvili2024searcher} examine how competition among searchers influences the allocation of value between a single validator and multiple searchers.
Wu et al. \cite{wu2024strategic, wu2024compete} investigate strategic behavior among builders in MEV-Boost auctions under the PBS framework, utilizing agent-based simulations and EGTA. They analyze various bidding strategies and latency conditions, finding that a decentralized builder market often incentivizes collusion, resulting in suboptimal auction outcomes. 
Braga et al.~\cite{braga2024mev} extend the analysis of MEV and PBS by proposing a dynamic protocol-level MEV sharing mechanism modeled as an adjustable extraction rate. They analyze its evolution and stability through dynamical systems theory, demonstrating that suitable adjustment intensities yield a stable equilibrium between users and block producers. Even under chaotic dynamics, the system maintains bounded deviations from equilibrium and ensures long-term market liveness, thus robustly guaranteeing MEV sharing.


\section{Model}\label{sec:model}

\begin{figure}[h]
	\centering	\includegraphics[width=0.5\textwidth, trim=0 100 150 40, clip]{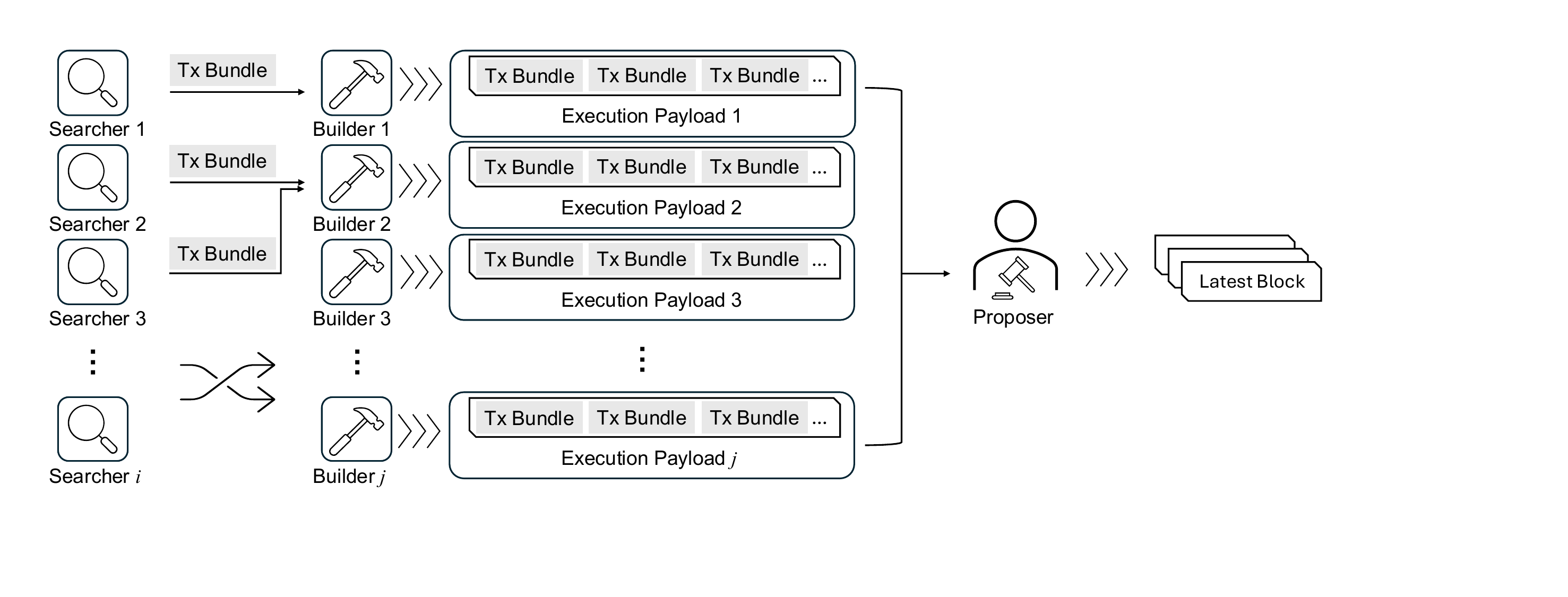}
	\caption{Sketch of the PBS ecosystem}\label{sketch}
\end{figure}

In the PBS ecosystem illustrated in Fig.~\ref{sketch}, specialized \textit{Private Order Flow} (POF) providers—known as searchers—send valuable bundles containing their own transactions and bids for inclusion directly to block builders via private channels, bypassing the public mempool to protect their strategies and arbitrage opportunities. Block builders aggregate these differentiated order flows and participate in the block building auction.

Let $\mathcal{A} = \{a_1,a_2,\cdots,a_n\}$, where $n>2$, represent a collection of agents participating in block building process. 
We represent the agents' index set by $\mathcal{N} = \{1,2,\cdots,n\}$.
Each agent $a_i$ is associated with a unique arbitrage opportunity defined as a bundle $\mathbf{b}_i$. 
The bundle must be included in a block $\mathcal{B}_j$ and land on the chain to effectively extract value.
A block is an ordered tuple $(\mathbf{b}_{(1)},\mathbf{b}_{(2)},\cdots,\mathbf{b}_{(m)})$ composed of bundles, where the index indicates the execution order of the bundle within the block.
Within a bundle, transactions are executed sequentially to capture profit opportunities in the market, but their outcomes may be influenced by the presence and ordering of other bundles within the same block. We signify the effective value of the bundle $\mathbf{b}_i$ in block $\mathcal{B}_j$ as $v(\mathbf{b}_i,\mathcal{B}_j)$.

Subsequently, we quantitatively characterize the impact of the execution order on the interaction among these bundles and their respective values. 
Consider a directed weighted graph $(\mathcal{G}(V,E),\Phi)$ with weight matrix $\Phi=(\varphi_{ij})_{n\times n}$, where the set of nodes $V=\{\mathbf{b}_1,\mathbf{b}_2,\cdots,\mathbf{b}_n\}$ represents the pending bundles, and $E$ denotes the set of directed edges indicating the exisitence of interactions between bundles. The weights of nodes, denoted as $v_i$, signifies the effective value of $\mathbf{b}_i$, the weights of edges are specified by $\Phi$. For any two bundles $\mathbf{b}_i$ and $\mathbf{b}_j$, if and only if they are independent, then $\varphi_{ij} = \varphi_{ji} = 0$, signifying no connection edge between them, and vice versa. If $\mathbf{b}_i$ and $\mathbf{b}_j$ are competitive, then $\varphi_{ij}<0$ and $\varphi_{ji}<0$,  and if they are altruistic, then $\varphi_{ij}>0$ and $\varphi_{ji}>0$.
Upon being added to a block $\mathcal{B}_j$, the effective value $v(\mathbf{b}_i,\mathcal{B}_j)$ is updated for once a bundle $\mathbf{b}_k$ executed prior over $\mathbf{b}_i$
\begin{align*}
v(\mathbf{b}_i,\mathcal{B}_j) \leftarrow 
v(\mathbf{b}_i,\mathcal{B}_j) + \varphi_{ik}\cdot v(\mathbf{b}_i,\mathcal{B}_j).
\end{align*}

\subsection{Agent Policies}
In our model, we transform the roles of searchers and builders into two strategies employed by all agents: bundle sharing and block building.\footnote{As reported by \cite{Titan}, searchers can employ a ``shotgun'' tactic, meaning they submit to more than four builders or even all known builders.
Such searcher submission preferences are presumed by \cite{capponi2024proposer} as contributing to the utility of a prototypical order flow provider. 
Our model relaxes the relevant assumptions by considering agents as risk-neutral profit seekers, whose behavior results from strategic co-evolution in a two-sided market.} 
Bundle sharing strategy involves a bidding vector $\mathbf{p}_i=(p_{i1},p_{i2},\cdots,p_{in})$ of agent $i$, where $p_{ij}:=p_{ij}(v(\mathbf{b}_i,\mathcal{B}_{j})) $ represents the bid agent $i$ is willing to pay agent $j$ for inclusion of bundle $\mathbf{b}_i$ in the block $\mathcal{B}_j$.
In addition, 
\cite{raun2023leveraging} found that the most commonly used strategy involves bribing a certain proportion of profit, which means $p_{ij}(v(\mathbf{b}_i,\mathcal{B}_{j})) = \beta_{ij}\cdot v(\mathbf{b}_i,\mathcal{B}_j), \beta_{ij}\in(0,1)$.
Even the top-earning searchers utilize this method due to its simplicity. It allows the bid to adjust linearly with profit, ensuring that the bid does not exceed the profit during the block building phase. 
Bundle building as a strategy represents a rebate ratio $\alpha_{j}\in[0,1)$ of agent $j$, where  means the proportion of builder surplus refunded to searcher.

If we assume the block building auction adopts a second-price auction (or is theoretically approximated as such), whereby block builders submit truthful bids and the winning builder pays the amount of the second-highest bid.
Define $\mathcal{B}^{*}$ as the winning block and $P_i$ as the auction payment made by builder $a_i$;  consequently, the payoff received by the block proposer is formulated as
\begin{align*}
\pi_{BP} = \max_{i\in\mathcal{N}}\{P_i\},
	\end{align*}
while payoff
of bundle sharing and block building strategies are derived as follows.
In the context of bundle sharing, the profit of searchers is obtained from two segments: the first segment pertains to the value retained from their own bundle upon landing on the chain, while the second segment consists of a fraction of the builders' surplus refunded according to their bids' contribution to the overall block value.

\begin{align*}
	 \pi_i(\beta_i)= 
	\begin{cases}
 (1-\beta_{ij})v(\mathbf{b}_i,\mathcal{B}_{j})  \\
 +\alpha_{j}\frac{p_{ij}}{\sum\limits_{k\in\mathcal{N}/\{j\}} p_{kj}}\left(\sum\limits_{k\in\mathcal{N}} p_{kj}-P_{j}\right), &  \mathbf{b}_i\in \mathcal{B}_j=\mathcal{B}^{*}  , \\
		0,  &  \mathbf{b}_i\notin\mathcal{B}_j=\mathcal{B}^{*}.
	\end{cases}
\end{align*}

For block building, builders' earnings constitute the surplus obtained after winning the block building auction, minus the portion returned to searchers.
\begin{equation*}
\pi_j(\alpha_j) = 
\begin{cases}
(1-\alpha_j)\left(\sum_{i\in\mathcal{N}}p_{ij} - P_j\right), & \mathcal{B}_j = \mathcal{B}^{*},\\
0, & \mathcal{B}_j\neq \mathcal{B}^{*}.
\end{cases}
\end{equation*}

In a one-sided market with one searcher and two builders, a rational searcher will disregard the builders’ rebate ratio signals and always choose to submit a bid of zero, leading the builders to in turn set the surplus refund to zero. This follows from the analysis of the Bayesian Nash equilibrium in the corresponding extensive game, the details of which are presented below.  Specifically, consider a setting with $|\mathcal{A}| = 3$, where agents $a_1$ and $a_2$ serve as builders and $a_3$ acts as the searcher. Due to severe conflict between the builders, with $\varphi_{1,2} = \varphi_{2,1} = -1$, they cannot share bundles with each other. We further assume that $a_3$ always adopts the bundle-sharing strategy. Each agent's bundle value $v_i$ is independently drawn from a known distribution $F_i$. Agents do not observe each other's private valuations, but the value distributions are common knowledge.  

The payoff functions $\pi_1$, $\pi_2$, and $\pi_3$ are defined according to the values of the corresponding variables and can be summarized as follows. 

For agent $a_1$, the payoff function is given by $\pi_1(v_2, v_3) = (1 - \alpha_{13}) [v_1 - v_2 + (\beta_{31} - \beta_{32}) v_3]$ if $v_1 + \beta_{31} v_3 \geq v_2 + \beta_{32} v_3$; otherwise, $\pi_1(v_2, v_3) = 0$. 

Similarly, for agent $a_2$, when $v_2 + \beta_{32} v_3 \geq v_1 + \beta_{31} v_3$, we have $\pi_2(v_1, v_3) = (1 - \alpha_{23}) [v_2 - v_1 + (\beta_{32} - \beta_{31}) v_3]$, and $\pi_2(v_1, v_3) = 0$ otherwise. 

For agent $a_3$, the payoff is $\pi_3(v_1, v_2) = (1 - \beta_{31}) v_3 + \alpha_{13} [v_1 - v_2 + (\beta_{31} - \beta_{32}) v_3]$ if $v_1 + \beta_{31} v_3 \geq v_2 + \beta_{32} v_3$, and $\pi_3(v_1, v_2) = (1 - \beta_{32}) v_3 + \alpha_{23} [v_2 - v_1 + (\beta_{32} - \beta_{31}) v_3]$ otherwise.  

Denote $\mathrm{\Delta} v = v_1 - v_2$ and $\mathrm{\Delta}\beta = \beta_{31} - \beta_{32}$. Then, the payoff function $\pi_3(\mathrm{\Delta} v)$ can be written as follows: when $\mathrm{\Delta} v \geq -\mathrm{\Delta}\beta\, v_3$, we have  
$\pi_3(\mathrm{\Delta} v) = v_3 - (\beta_{32} + \mathrm{\Delta}\beta)\, v_3 + \alpha_{13} (\mathrm{\Delta} v + \mathrm{\Delta}\beta\, v_3)$;  
otherwise, i.e., when $\mathrm{\Delta} v < -\mathrm{\Delta}\beta\, v_3$, the payoff becomes  
$\pi_3(\mathrm{\Delta} v) = v_3 - \beta_{32} v_3 + \alpha_{23} \big( -\mathrm{\Delta} v - \mathrm{\Delta}\beta\, v_3 \big)$.

Under the assumptions that $v_1 \sim \text{Exp}(\lambda_1) $ and $v_2\sim \text{Exp}(\lambda_2) $,
we can deduce
$\mathrm{\Delta} v$ follows a double exponential distribution (Laplace distribution) with 
probability density function
\begin{equation*}
f_{\mathrm{\Delta} v}(x) =
\begin{cases}
\frac{\lambda_1 \lambda_2}{\lambda_1+\lambda_2}e^{-\lambda_1 x}, & x\geq0,\\
\frac{\lambda_1 \lambda_2}{\lambda_1+\lambda_2}e^{\lambda_2 x}, & x<0.
\end{cases}
\end{equation*}
Accordingly, the expected payoff $\mathbb{E}[\pi_3]$ can be expressed as  
\begin{align*}
&\mathbb{E}[\pi_3]\\
=& \int_{-\mathrm{\Delta}\beta v_3}^{+\infty}[v_3-(\beta_{32}+\mathrm{\Delta}\beta)v_3+\alpha_{13}(x+\mathrm{\Delta}\beta v_3)]f_{\mathrm{\Delta} v}(x)\mathrm{d}x\\
& + \int_{-\infty}^{-\mathrm{\Delta}\beta v_3}[v_3-\beta_{32}v_3+\alpha_{23}(-x-\mathrm{\Delta}\beta v_3)]f_{\mathrm{\Delta} v}(x)\mathrm{d}x
\end{align*}
It is evident that when $\mathrm{\Delta}\beta\geq0$ is given, setting $\beta_{32}=0$ is a dominant strategy. This aligns with economic intuition, as the factor affecting the builder's winning probability is $\mathrm{\Delta}\beta$, and setting $\beta_{32}=0$ minimizes cost. 
By substituting $f_{\mathrm{\Delta} v}(x)$ and denoting $K=\frac{\lambda_1\lambda_2}{\lambda_1+\lambda_2}$, $C_1 =v_3-\mathrm{\Delta}\beta v_3+\alpha_{13}\mathrm{\Delta}\beta v_3$, $C_2= v_3-\alpha_{23}\mathrm{\Delta}\beta v_3$, we can obtain
\begin{align*}
& \mathbb{E}[\pi_3]\\
= &
 \int_{0}^{+\infty}[v_3-\mathrm{\Delta}\beta v_3+\alpha_{13}(x+\mathrm{\Delta}\beta v_3)]\frac{\lambda_1 \lambda_2}{\lambda_1+\lambda_2}e^{-\lambda_1 x}\mathrm{d}x\\
& + \int_{-\mathrm{\Delta}\beta v_3}^{0}[v_3-\mathrm{\Delta}\beta v_3+\alpha_{13}(x+\mathrm{\Delta}\beta v_3)]\frac{\lambda_1 \lambda_2}{\lambda_1+\lambda_2}e^{\lambda_2 x}\mathrm{d}x\\
& + \int_{-\infty}^{-\mathrm{\Delta}\beta v_3}[v_3+\alpha_{23}(-x-\mathrm{\Delta}\beta v_3)]\frac{\lambda_1 \lambda_2}{\lambda_1+\lambda_2}e^{\lambda_2 x}\mathrm{d}x.
\end{align*}
Computing the derivative of $\mathbb{E}[\pi_3]$ with respect to $\mathrm{\Delta}\beta$, we have
\begin{align*}  
& \frac{\mathrm{d}\mathbb{E}[\pi_3]}{\mathrm{d}\mathrm{\Delta\beta}} \\
=\, & \frac{\lambda_1 \lambda_2}{\lambda_1 + \lambda_2} \Bigg(  
    \frac{v_3(\alpha_{13} - 1)}{\lambda_1}   
    + \frac{v_3(\alpha_{13} - 1)}{\lambda_2}  
      \left[1 - e^{-\lambda_2 \mathrm{\Delta}\beta v_3}\right] \\
& \hspace{2.8cm}  
    - v_3^2\, \mathrm{\Delta}\beta\, e^{-\lambda_2 \mathrm{\Delta}\beta v_3}  
    - \frac{\alpha_{23} v_3\, e^{-\lambda_2 \mathrm{\Delta}\beta v_3}}{\lambda_2}  
    \Bigg) \\
<\, & 0.  
\end{align*}  
This implies that the searcher should adopt a $\mathrm{\Delta}\beta=0$ strategy, meaning uniform bidding to all builders. However, in a two-sided market that incorporates randomly interacting bundles, the situation becomes more complex and challenging to analyze directly, which will be the main focus of our subsequent discussion.

\subsection{Modeling of Co-evolution}
\label{MCE}
\begin{figure}[!htbp]
    \centering
\includegraphics[width=0.45\textwidth, trim=30 0 30 0, clip]{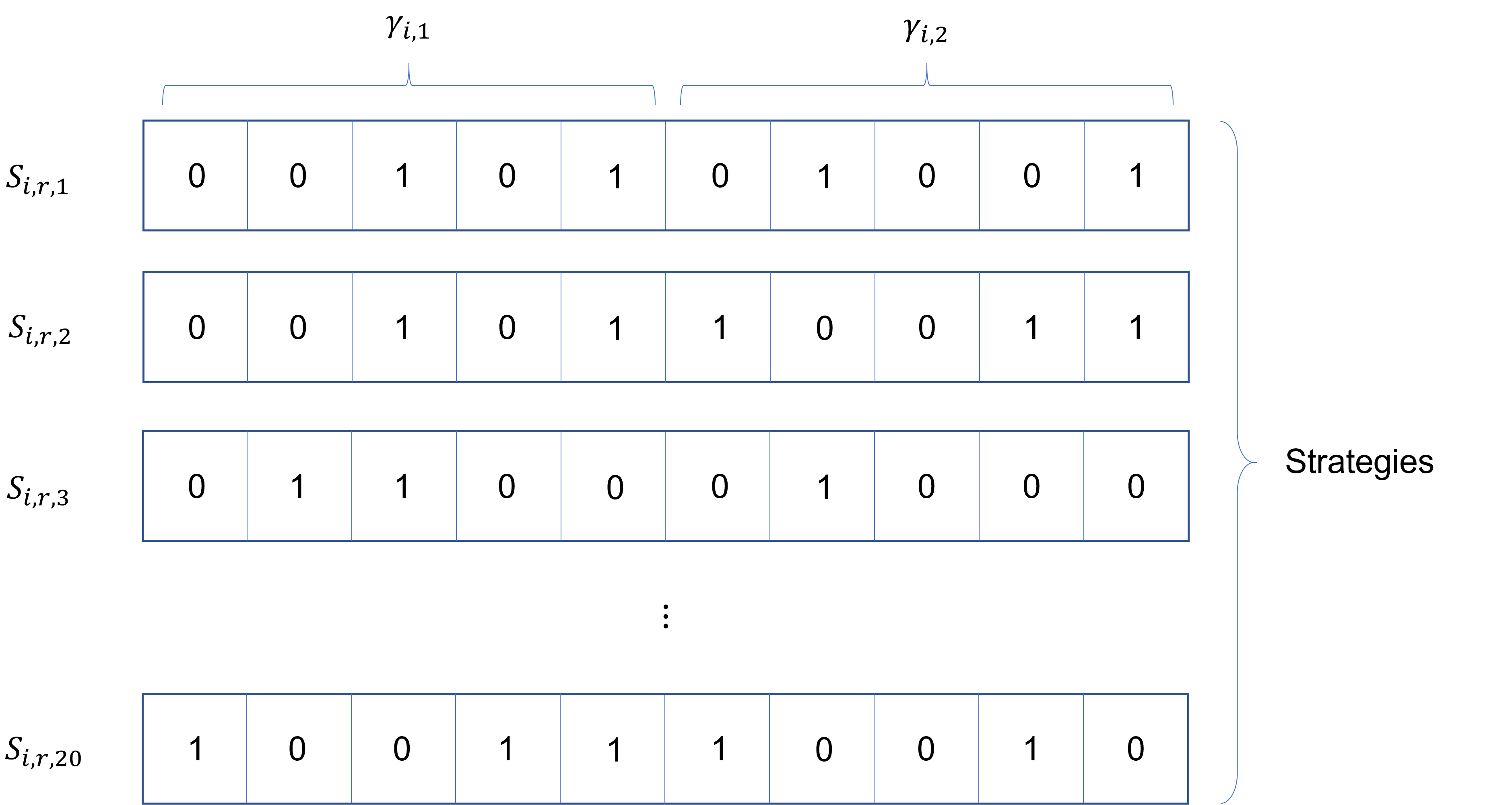}
    \caption{Searcher $i$'s strategy population in the $r$th generation}
    \label{strategy}
\end{figure}

In our modeling, we employ genetic algorithms to capture the co-evolution process of agents. Each agent begins with a randomly generated population of 20 strategies. The block building strategy comprises a 5-digit binary sequence referred to as the ``chromosome", indicating the relevant rebate ratio $\alpha_j$.
As shown in Fig. \ref{strategy}, for bundle sharing, $S_{i,r,k}$ signifies the $k$th strategy of searcher $i$ in the $r$th generation. Each 5-digit segment within a chromosome corresponds to a parameter value that regulates the bidding behaviors of bundle sharing strategy.
The bid ratio is computed using a modified sigmoid function 
\begin{align*}
\beta_{ij} = \left( \frac{1}{1+\gamma_{i,1}^{-\alpha_{j}}}\right)^{\gamma_{i,2}},
\end{align*}
where $\gamma_{i,1}\in[1,5]$ 
measures the sensitivity of searchers' bidding inclinations towards builders with elevated rebate ratio, whereas $\gamma_{i,2}\in[0,4]$ determines the overall scale of the bid ratio. The adoption of the modified sigmoid function ensures that the bid ratio $\beta_{ij}$ remains within the interval $[0,1]$ and 
is monotonically increasing in relation to $\alpha_j$.
The largest decimal number that a 5-digit binary number can represent is $31$.
Taking $S_{i,r,1}$ ``0010101001" as an example, the first 5-digit binary segment ``00101" can be translated to the decimal number number $1\cdot 2^2 + 1\cdot 2^0 = 5$.
Thus, the parameter $\gamma_{i,1}$ can be derived from ``00101" as $1+(5/31)\cdot(5-1) = 21/31$. Similarly, the second 5-digit binary ``01001" corresponds to the parameter  $\gamma_{i,2} = 0 + (9/31)\cdot(4-0) = 36/31$ suggesting that $S_{i,r,1}$ ``0010101001" corresponds to the parameter pair $(\gamma_{i,1},\gamma_{i,2}) = (21/31,36/31)$. The rebate ratio $\alpha_j$ for block building is directly obtained through a mapping that converts a 5-digit binary number into a decimal value within the interval $[0,1]$.

Agent-based simulation, based on reinforcement learning principles, involves each agent iterating the through following steps continuously:

     \textit{Step 1}: A private value is randomly generated for each agent, representing an opportunity available to them, alongside the random creation of an interaction matrix. In our subsequent experiments, the private values adhere to an exponential distribution with a default parameter of 10 (mean value of 0.1).\footnote{Given a fixed mean (or rate parameter), the exponential distribution is the exclusive distribution that fulfills the maximum entropy requirement, thereby introducing the fewest extra assumptions.} For simplicity, the interaction matrix is constructed according to a two-point distribution $\mathbb{P}(\varphi_{ij}=\varphi_{ji}=-1) = p_{C} $ and  $\mathbb{P}(\varphi_{ij}=\varphi_{ji}=0) = 1-p_{C}$. This implies that for any two agents, they completely conflict with a probability of $p_{C}$, or they operate independently with a probability of $1-p_{C}$.\footnote{As pointed in \cite{mamageishvili2024searcher}, order flow providers possessing complementary flows are incentivized to integrate in order to collaboratively capture greater value, which consequently leads to a form of interaction between bundles that is predominantly characterized by conflict.}
  
     \textit{Step 2}: Agent selects a strategy from the strategy repository using a roulette-wheel selection method based on the fitness of the strategies $f(S_{i,r,k})$, 
    \begin{align*}
    \mathbb{P}(S_{i,r,k} ~\text{is selected}) = \frac{e^{f(S_{i,r,k})/T}}{\sum_{k}e^{f(S_{i,r,k})/T}},
    \end{align*}
    where $T$ controls the trade-off between exploration and exploitation; through experimentation, we set $T=2$.  
At the beginning of each simulation, agents are randomly assigned a population of 20 strategies. Each agent maps its selected strategy (chromosome) to executable actions. Specifically, builder agents determine their rebate ratio $\alpha_j$, and searcher agents establish their bid ratio $\bm{\beta}_i$.

    \textit{Step 3}: Based on the individual strategies of the agents, the simulation of the block building process unfolds: searchers send bundles to builders, who utilize the bundles received from searchers along with their own bundles to construct the valuable blocks. We employed a greedy algorithm (Algorithm \ref{greedy_algorithm}) to simulate the merging of bundles into a block.\footnote{This study focuses on conditions of sufficient block space, wherein the total quantity of bundles does not exceed the capacity of the block. Nevertheless, our findings can be readily adjusted to address cases with capacity limitations on blocks, where a block can accommodate a maximum of $L$ bundles.} Builders participate in the block building auction to calculate the final rewards received by both builders and searchers from the winning block.
    Update the strategy fitness based on the strategies selected by the agents in this round.
    \begin{align*}
    f(S_{i,r+1,k}) = (1-\eta_i)\cdot f(S_{i,r,k}) + \eta_i\cdot\pi_{i,r}, 
    \end{align*}
    where $\eta_i$ regulates the balance between the emphasis on historical and recent performance of the strategies, we set a default value of $0.5$.

    \begin{algorithm}[htbp]
	\caption{Greedy algorithm for block building (bundles merging) based on bids}\label{greedy_algorithm}
	\begin{algorithmic}[1]
		\Require \\A list of pending bundles $\mathbb{B} = \{\mathbf{b}_1,  \mathbf{b}_2, \cdots,  \mathbf{b}_m\}$ \\
		An empty list $\mathcal{B}$ for storing selected bundles with maximum size $L$ 
		\Ensure 
		\While{the length of $\mathcal{B}$ is less than $L$ and $\mathbb{B}$ is not empty}
		\State Sort $\mathbb{B}$ such that bundles with effective value greater than 0 and higher effective priority fee (bid) come first
		\State Remove the first bundle from $\mathbb{B}$ and assign it to $\mathbf{b}^{*}$
		\For{each bundle $\mathbf{b}_i$ in $\mathbb{B}$}
		\State Calculate and update effective value of $\mathbf{b}_i$ 
		\EndFor
		\If{effective value of $\mathbf{b}^{*}$ is not greater than 0}
		\State Exit the loop as no more effective bundles are left
		\EndIf
		\State Add $\mathbf{b}^{*}$ to $\mathcal{B}$
		\EndWhile
	\end{algorithmic}
\end{algorithm}
    
    \textit{Step 4}: Agents optimize their strategy repository through genetic algorithms with a trigger probability of $0.01$. Initially, a subset of strategies is removed from the repository based on a specified elimination ratio (configured as $0.5$ in the simulation). Grounded in the ``chromosomes'', genetic algorithms integrate selection, crossover, and mutation processes until the strategy repository is replenished back to its upper limit of $20$, as detailed below.
    \begin{itemize}
\item \textit{Selection}:
Agents randomly select two parent strategies from the remaining strategies utilizing the roulette-wheel selection method.
\item \textit{Crossover}:
Two parent strategies are combined to create two offspring strategies by exchanging segments of the ``chromosome" representations of the parent strategies. 
The fitness of offspring strategies is calculated as the mean of the fitness values of their parent strategies.
\item \textit{Mutation}:
For every bit in the binary chromosomes of each offspring produced, a flip ($0\rightarrow1$ and $1\rightarrow0$) is performed with a specified mutation rate of $0.01$.
\end{itemize}

\section{Results}\label{sec:results}
In this section, we present the results of agent-based simulations. We first analyze the co-evolution of agents using genetic algorithms, then assess how bundle interactions affect system dynamics and payoffs. Finally, we examine and predict participants’ role selection in the block building market using EGTA.

\subsection{Co-evolution of Agents' Strategies}\label{CE}

\begin{figure*}[!htbp]  
    \centering  
    \begin{subfigure}[b]{0.32\textwidth}  
        \centering  
        \includegraphics[width=\linewidth]{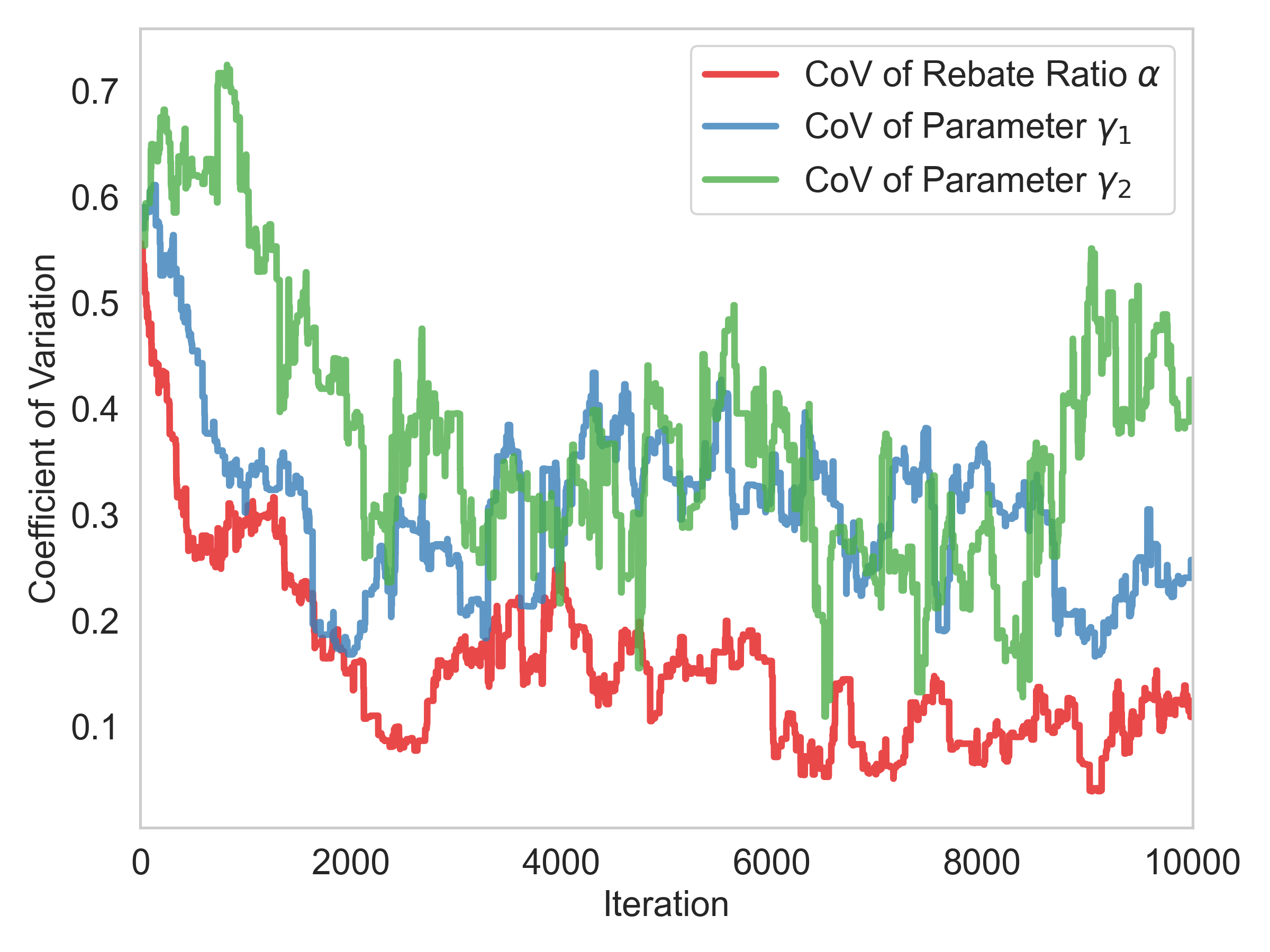} 
        \caption{Evolution of CoV}  
        \label{fig_CoV}  
    \end{subfigure}  
    \hfill  
    \begin{subfigure}[b]{0.32\textwidth}  
        \centering  
    \includegraphics[width=\linewidth]{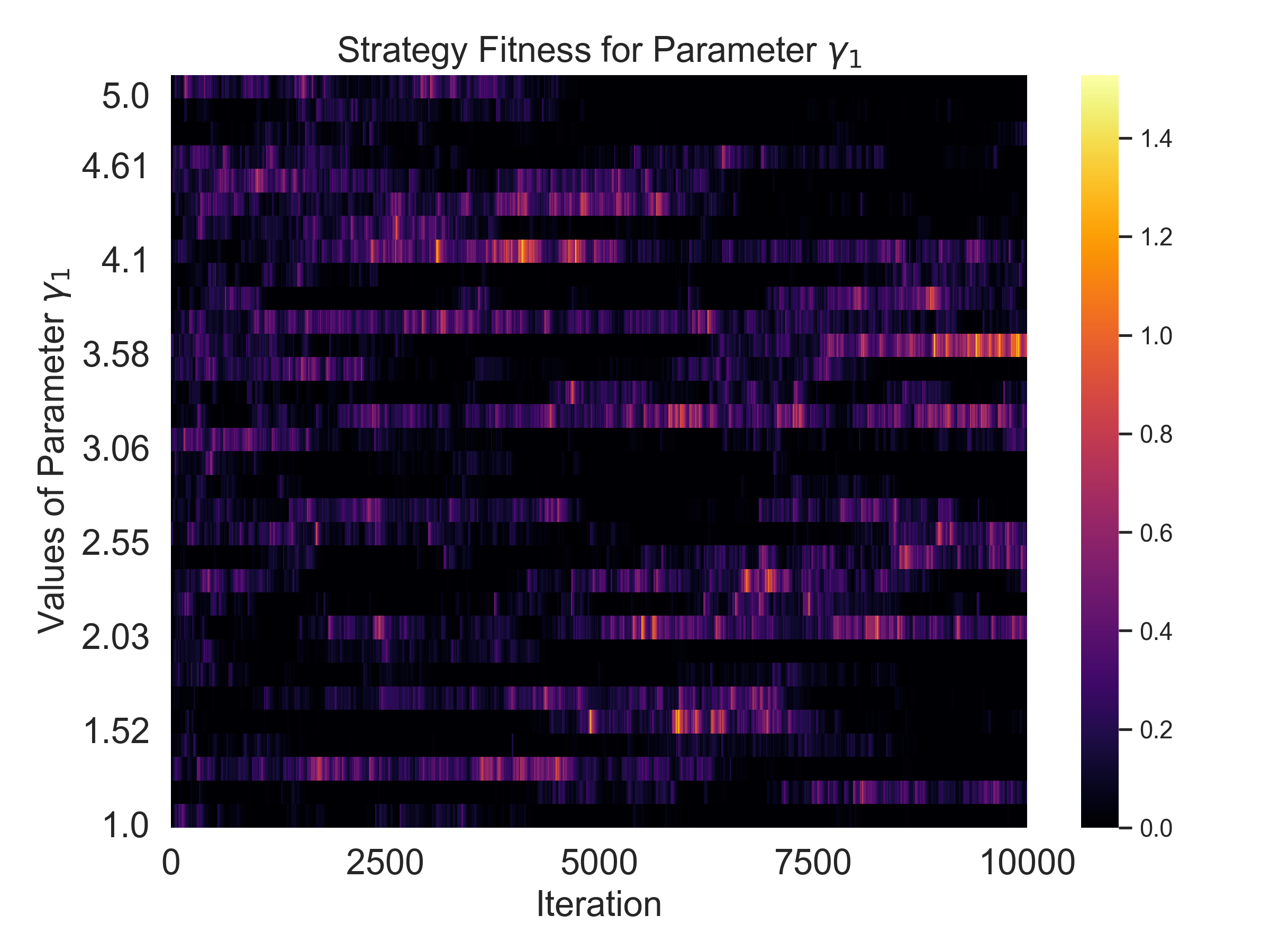}  
        \caption{Evolution of fitness for $\gamma_1$}  
        \label{gamma_1}  
    \end{subfigure}  
    \hfill  
    \begin{subfigure}[b]{0.32\textwidth}  
        \centering  
        \includegraphics[width=\linewidth]{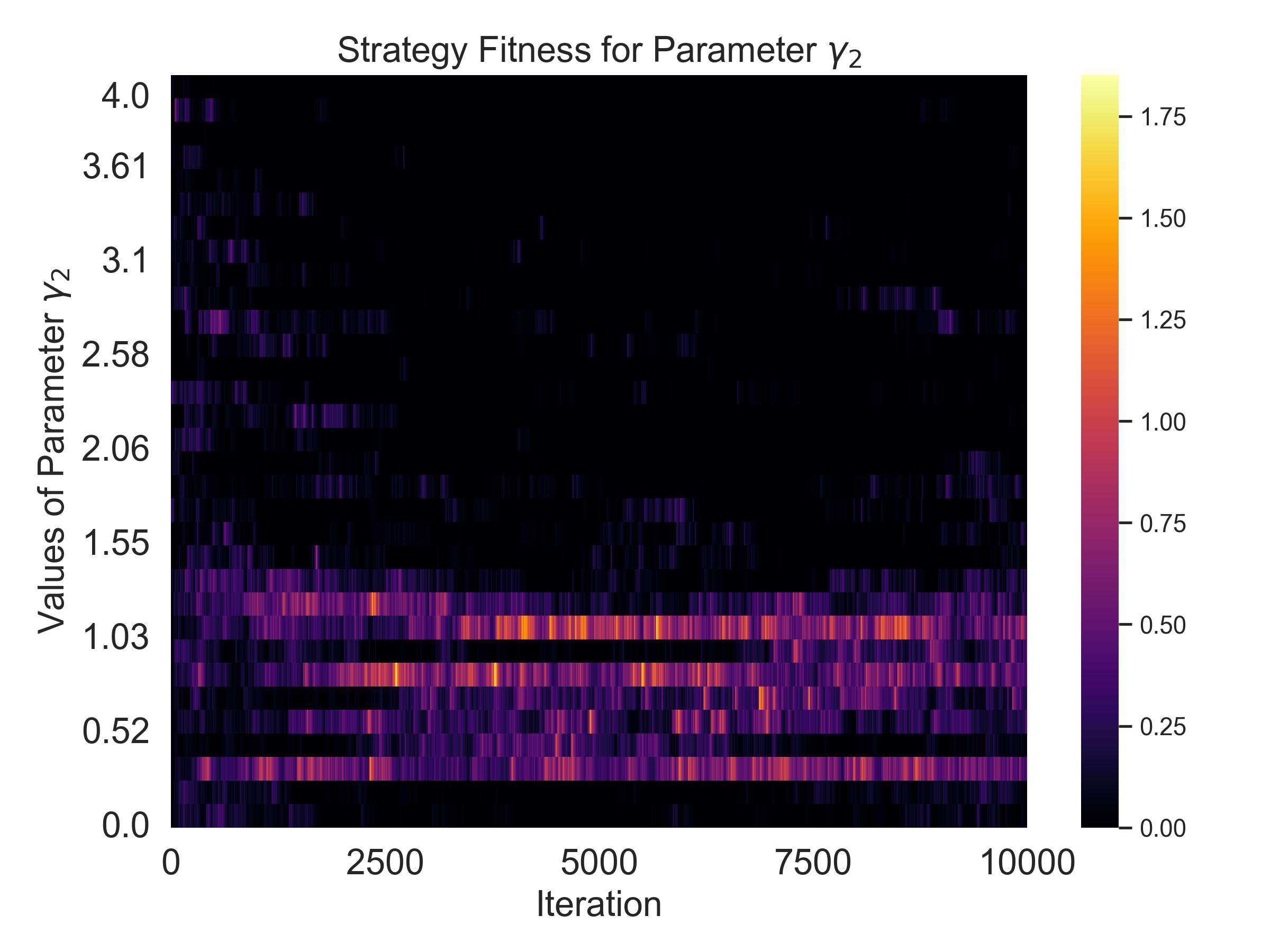}  
        \caption{Evolution of fitness for $\gamma_2$}  
        \label{gamma_2}  
    \end{subfigure}  
    
    \vspace{8pt} 
    
    \begin{subfigure}[b]{0.32\textwidth}  
        \centering  
        \includegraphics[width=\linewidth]{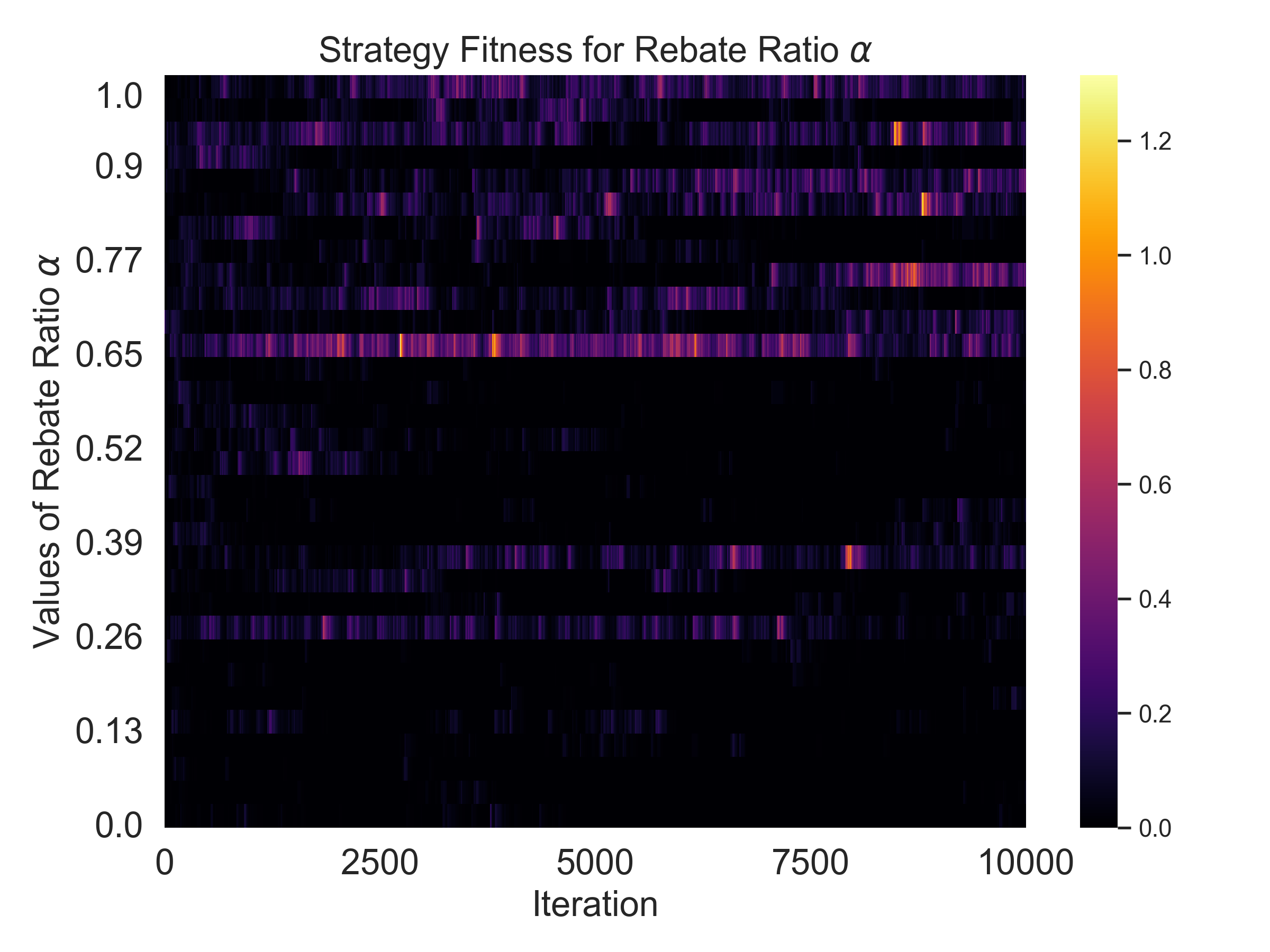}  
        \caption{Evolution of fitness for $\alpha$}  
        \label{alpha}  
    \end{subfigure}  
    \hfill  
    \begin{subfigure}[b]{0.32\textwidth}  
        \centering  
        \includegraphics[width=\linewidth]{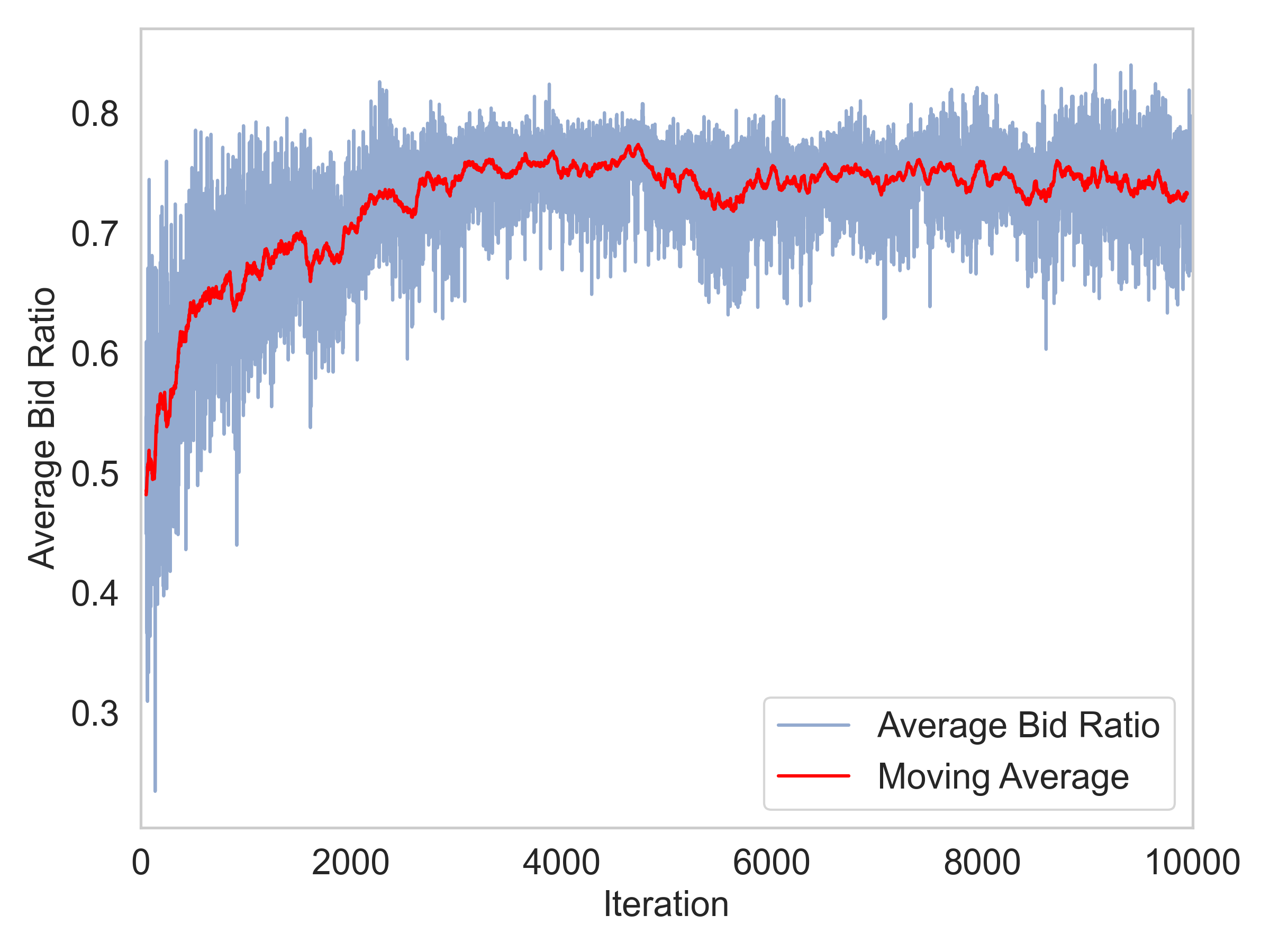}  
        \caption{Evolution of average bid ratio}  
        \label{bid_ratio}  
    \end{subfigure}  
    \hfill  
    \begin{subfigure}[b]{0.32\textwidth}  
        \centering  
        \includegraphics[width=\linewidth]{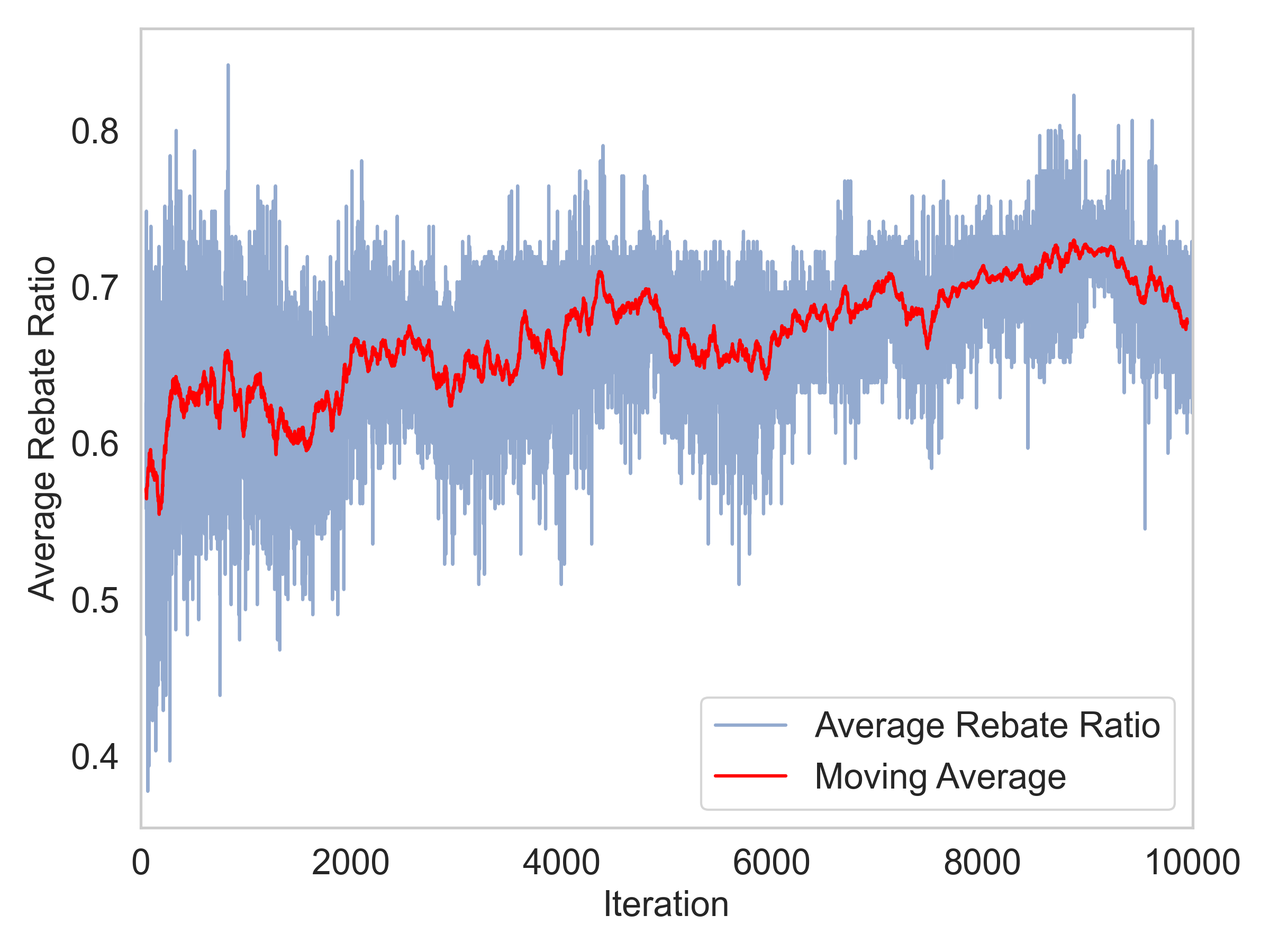}  
        \caption{Evolution of average rebate ratio}  
        \label{rebate_ratio}  
    \end{subfigure}  

    \caption{Co-evolution of agent strategies}  
    \label{evolution}  
\end{figure*}  

In this subsection, we simulate a model with 10 builders, 10 searchers, and a conflict probability of $p_C = 0.8$ over 10{,}000 blocks.

First, we assess the convergence of agent strategies by monitoring the coefficient of variation (CoV) for key strategy parameters over successive iterations. For each builder, 20 strategies for the rebate ratio~$\alpha$ are maintained per iteration; each strategy, represented as a 5-bit binary encoding, corresponds to a discrete integer in $[0, 31]$. The CoV for these values is computed as the standard deviation divided by the mean, with lower CoV values indicating greater consensus among strategies. Similarly, for searchers, we track the CoV of the behavioral parameters~$\gamma_1$ and~$\gamma_2$. Since all agents following the same meta-strategy are homogeneous, we report the average CoV for builders’ rebate ratios and for the searchers’ parameters across all agents in each iteration. Fig.~\ref{fig_CoV} demonstrates a clear downward trend in the CoV of $\alpha$, $\gamma_1$, and $\gamma_2$, declining from $0.56$, $0.57$, $0.56$ to $0.12$, $0.24$, and $0.43$ over 10,000 iterations, respectively. These results indicate that agents’ strategies generally converge over time.

To examine the evolutionary process of agents' behaviors, we calculated the cumulative fitness of all strategies within the agents' strategy repository for each iteration.
Figs. \ref{gamma_1}, \ref{gamma_2} and \ref{alpha} depict the evolution of the cumulative fitness of $\gamma_1$, $\gamma_2$ and $\alpha$, respectively. 
The fitness reflects the performance of the strategy and the agents' propensity to utilize it.
On one hand, we observe that only a small subset of strategies achieve high fitness (lighter shades in the figure), clustering around certain values, while most strategies have near-zero fitness, consistent with the overall decline in CoV. However, these high-fitness strategies are not entirely concentrated, indicating some diversity in agent behaviors.
Overall, searchers display differentiated bidding: they tend to offer higher bid ratios to builders with higher announced rebate ratios (Fig.~\ref{gamma_1}). They also generally choose smaller values of $\gamma_2$ to increase their overall bid ratios, reflecting intensified competition under $p_C=0.8$ (Fig.~\ref{gamma_2}). For builders, Fig.~\ref{alpha} shows that several extremely high rebate ratios correspond to high fitness, suggesting fierce competition and relatively low profitability for builders. We analyze agent profitability in detail in the following subsections. 

To provide a clearer illustration of the overall evolution of rebate ratios and bid ratios among the two types of agents within the system, we calculated the average values of bid ratios (for searchers) and rebate ratios (for builders) based on the actual actions taken in each iteration, as depicted in Figs. \ref{bid_ratio} and \ref{rebate_ratio}, respectively. 
The calculation of the moving average reveals a sustained upward trend in the average bid ratio, as shown in Fig. \ref{bid_ratio}, alongside the convergence of the average rebate ratio toward approximately 0.7, as indicated in Fig. \ref{rebate_ratio}.

Overall, our agent-based simulation demonstrates that agents successfully adapt to the PBS environment by autonomously learning strategies through reinforcement learning. This co-evolutionary process leads to the convergence of strategies, while preserving some degree of diversity.

\subsection{Impact of Bundle Interactions}\label{BI}
\begin{figure*}[!htbp]  
	\centering  
    \begin{minipage}[c]{0.19\textwidth}  
		\centering  
		\includegraphics[width=\textwidth]{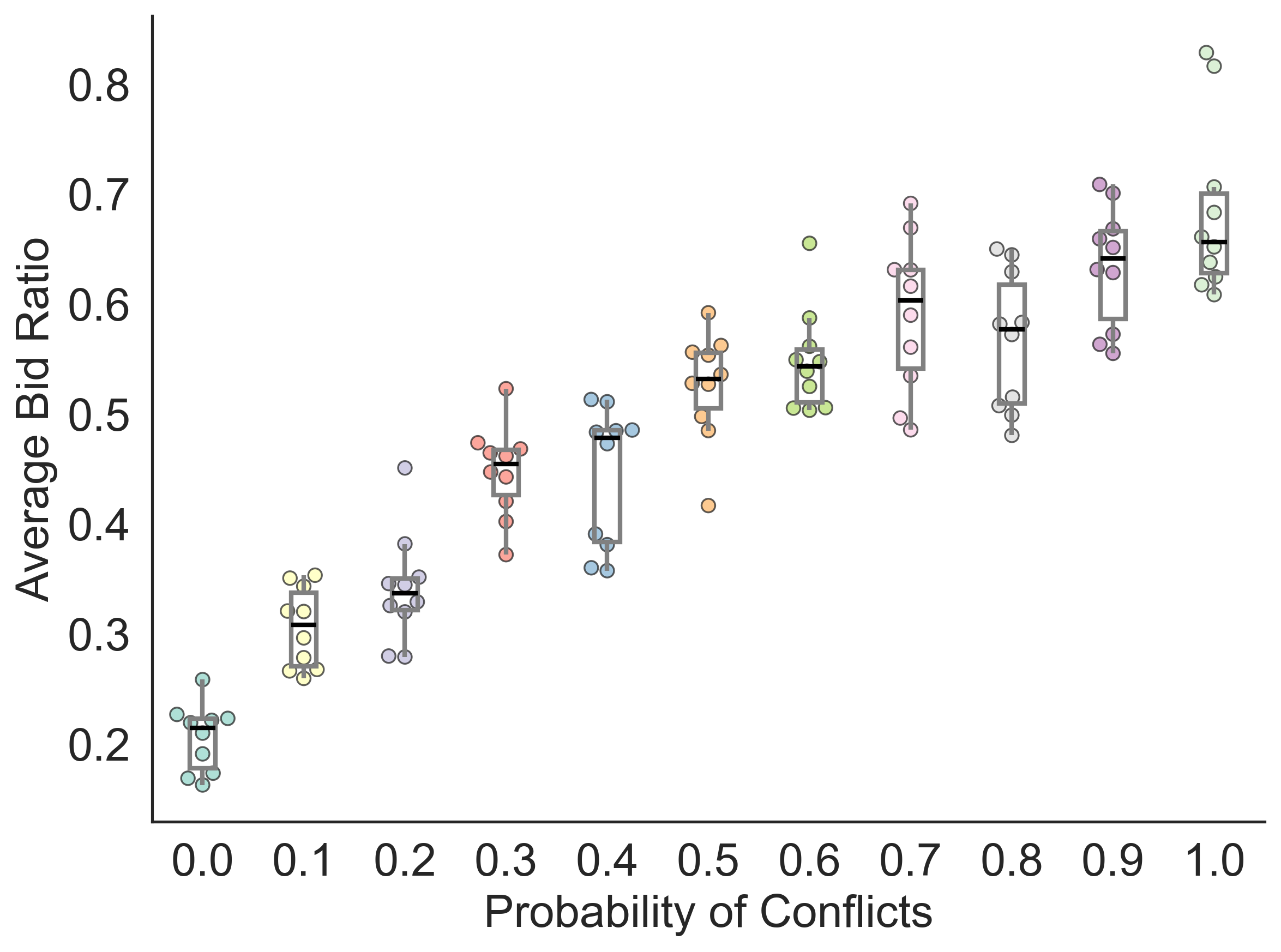}  
		\subcaption{}   
		\label{2_average_bid_ratio}  
	\end{minipage}    
	\begin{minipage}[c]{0.19\textwidth}  
		\centering  
		\includegraphics[width=\textwidth]{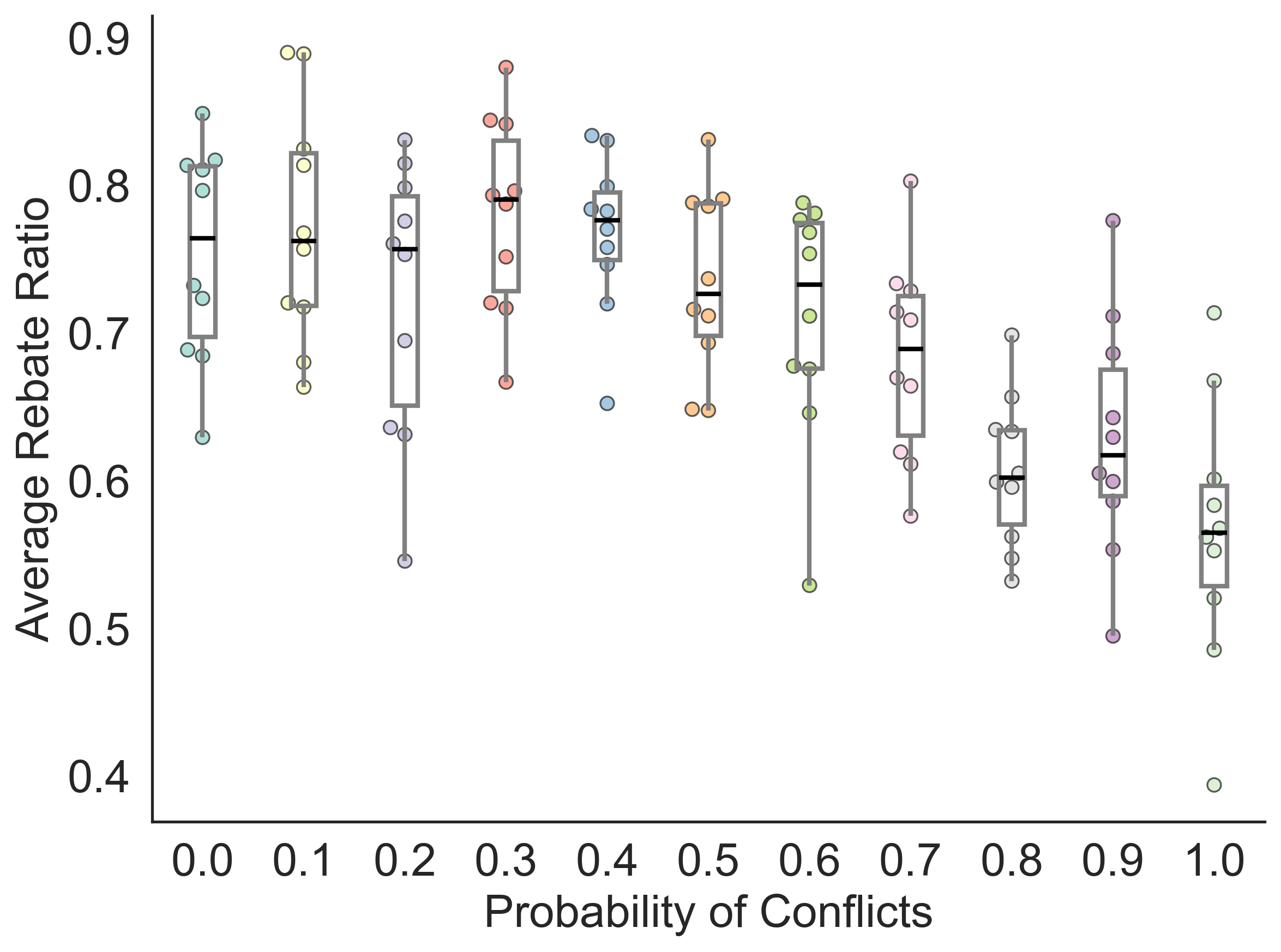}  
		\subcaption{}    
		\label{2_average_rebate_ratio}  
	\end{minipage}  
	\begin{minipage}[c]{0.19\textwidth}  
		\centering  
		\includegraphics[width=\textwidth]{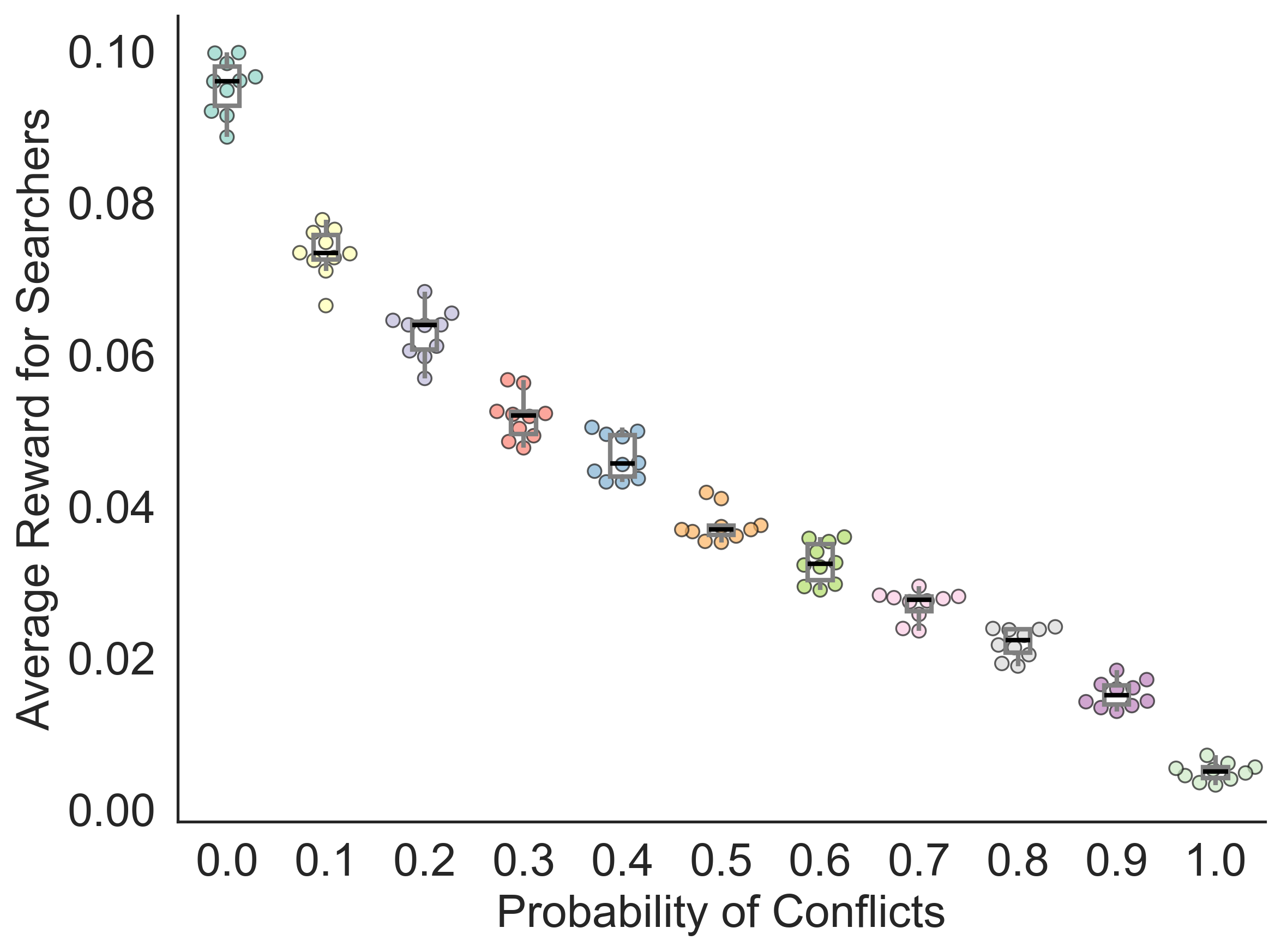}  
		\subcaption{}  
		\label{2_average_reward_searchers}  
	\end{minipage}  
	\begin{minipage}[c]{0.19\textwidth}  
		\centering  
		\includegraphics[width=\textwidth]{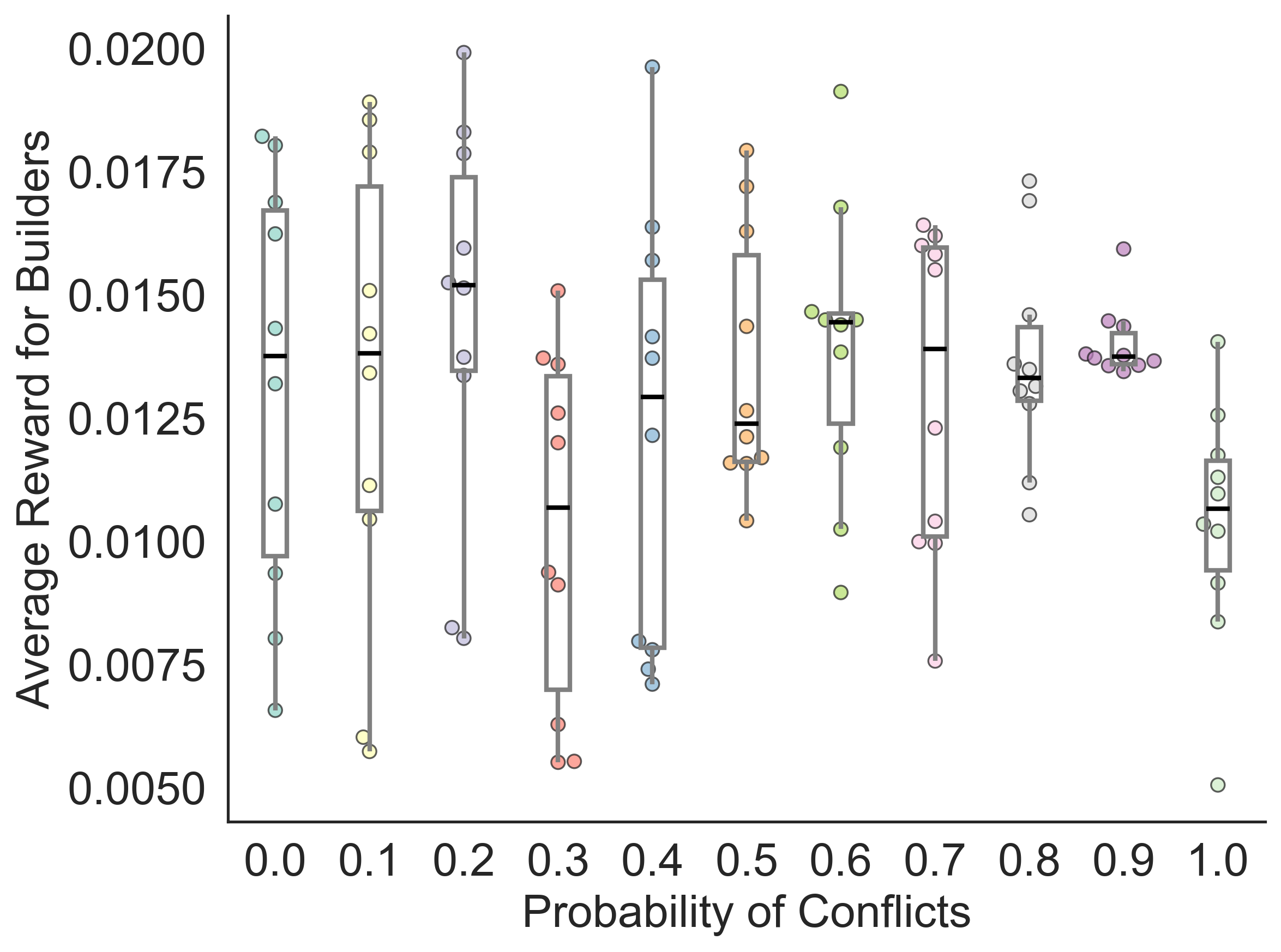}  
		\subcaption{}  
		\label{2_average_reward_builders}  
	\end{minipage}    
	\begin{minipage}[c]{0.19\textwidth}  
		\centering  
		\includegraphics[width=\textwidth]{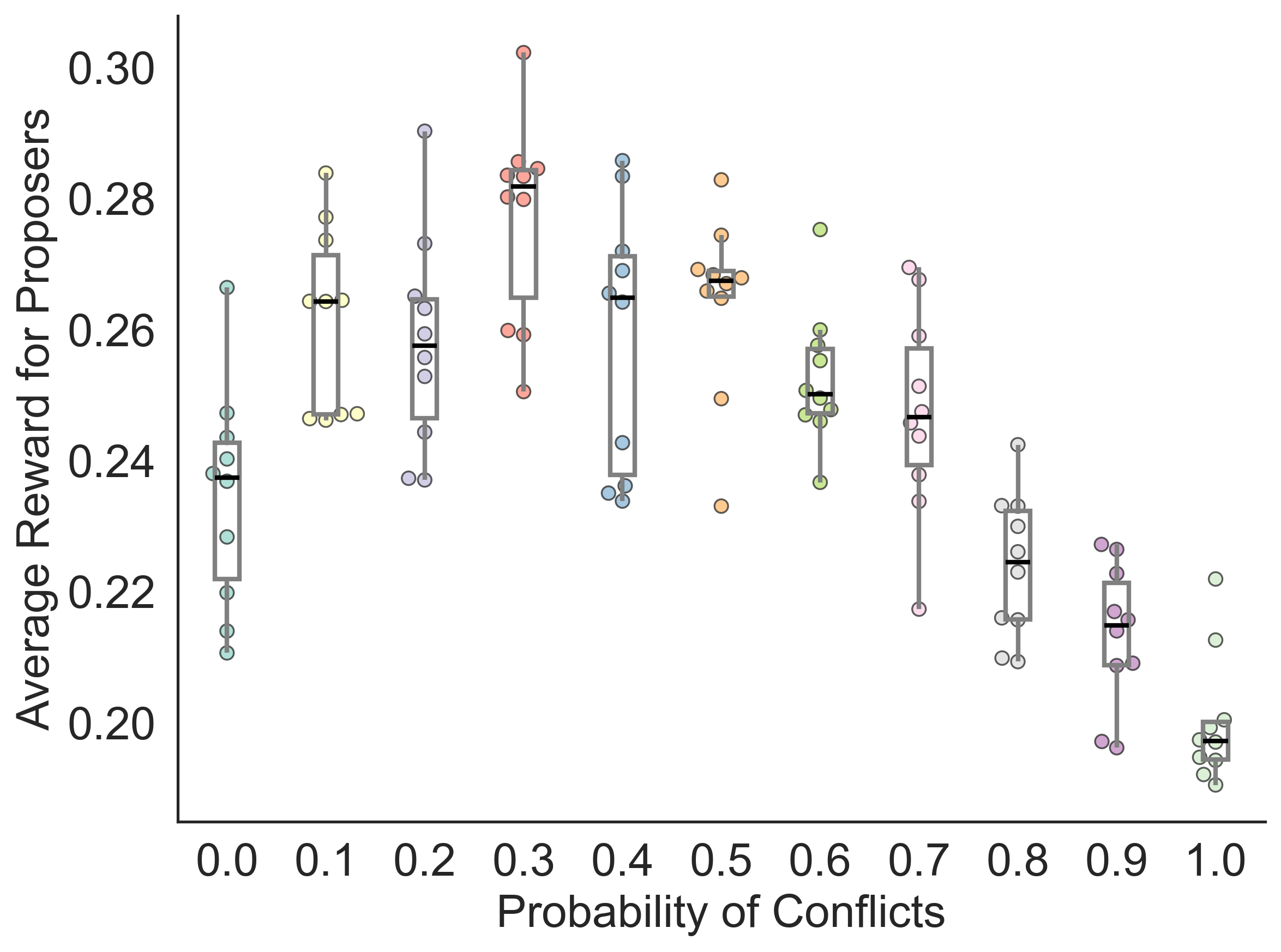}  
		\subcaption{}  
		\label{2_average_reward_proposers}  
	\end{minipage}  
    \caption{Impact of conflict probability} 
    \label{p}  
\end{figure*}

In this subsection, we focus on the impact of bundle interactions on the evolutionary outcomes of the system.
During the simulation process, we simplify the interactions among bundles to include conflicts that arise between any two bundles with a specified probability. By performing 10 repetitions for each designated conflict probability within a system comprising 5 builders and 5 searchers, we statistically analyze the average bid ratio of searchers, the average rebate ratio of builders, and the profits of searchers and builders' proposers after the system reaches a state of stability, yielding the results depicted in Fig.~\ref{p}.

According to Fig. \ref{p}, it is evident that the average bid ratio and average reward results of searchers exhibit less volatility compared to those of builders.
For searchers, an increase in the probability of conflicts significantly intensifies competition among them, elevating the overall level of their bid ratios while concurrently reducing their average returns, as depicted in Figs.
\ref{2_average_bid_ratio} and \ref{2_average_reward_searchers}. The impact of the probability of conflicts on builders is relatively minor; however, it exhibits a more complex non-monotonic pattern. When the probability of conflicts is low, an increase in this probability encourages builders to raise their rebate ratios, a result of co-evolution: searchers correspondingly increase their bid ratios, thereby providing more valuable bundles to builders. In turn, builders enhance their rebate ratios to capitalize on the differentiated order flow from searchers. However, once the probability of conflicts surpasses a certain threshold (0.3 in Fig. \ref{2_average_rebate_ratio}), more frequent bundle conflicts undermine the overall value of the searchers' bundles, despite the increase in their bid ratios. At this point, builders become more reliant on the value of their own bundles, and utilizing a lower rebate ratio allows them to retain a greater share of surplus.
From Fig. \ref{2_average_rebate_ratio}, we can see that the probability of conflicts has a minimal impact on builders' earnings. Conversely, the average reward for proposers exhibits a distinct pattern of initially increasing and then decreasing as the probability of conflicts rises, as shown in Fig. \ref{2_average_reward_proposers}. An increasing probability of conflicts compels searchers to adopt higher bid ratios; however, excessively high probabilities of conflicts lead to a reduction in the overall MEV of the block. Under the current settings, a probability of conflicts $p_C$ of 0.3 maximizes MEV.

\subsection{Empirical Game-Theoretic Analysis}\label{egta}
In this subsection, we treat agents' choices between bundle sharing and block building as two meta-strategies and analyze them using EGTA. We construct a meta-game with 10 agents, generating a heuristic payoff table $(N, U)$~\cite{tuyls2018generalised}, where $N_i$ specifies the distribution of agents across the two strategies ($N_{i1}$ and $N_{i2}$ represent the numbers of agents adopting block building and bundle sharing, respectively), and $U_i$ denotes the corresponding average payoffs derived from repeated agent-based simulations.
Based on the heuristic payoff table, we employ the $\alpha$-Rank algorithm \cite{omidshafiei2019alpha} to establish the Markov transition matrix for the two categories of strategies and calculate the stationary distribution of this transition matrix, which serves as a a dynamic solution for the meta-game to evaluate and rank the meta-strategies.
Furthermore, we calculate the stationary distribution under varying probabilities of conflict and conduct a sweep over the ranking intensity alpha as noted in \cite{omidshafiei2019alpha}, ranging from $0.1$ to $100$. An adequately large ranking intensity guarantees that $\alpha$-Rank maintains the ranking of strategies most closely aligned with the Markov-Conley chains solution concept.

\begin{figure*}[!htbp]
	\centering
\includegraphics[width=1\textwidth]{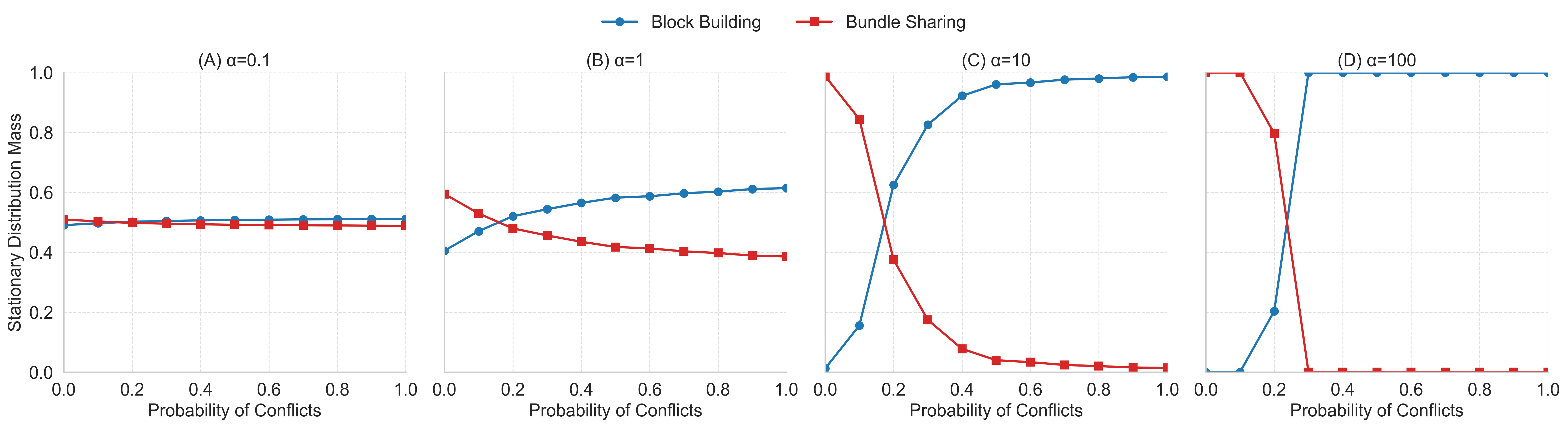}	
\caption{Stationary distribution of block building and bundle sharing}\label{rank}
\end{figure*}
Fig. \ref{rank} clearly demonstrates that an increase in the probability of conflicts leads agents to prefer the block building strategy, which entails building blocks to engage in competition within the block building auction, instead of resorting to the bundle sharing strategy where they share their bundles with builders to evade fierce competition. This result aligns well with intuitive reasoning.
In particular, we can identify the critical probability of conflicts at which agents' strategies transition, which is approximately $0.2$.

Our results show that when the probability of bundle conflicts is low, searchers capture the majority of value within the bundles, and builders must actively participate in block-building auctions to extract additional profits. As the likelihood of conflicts increases, however, competition among searchers intensifies, prompting higher bid ratios to ensure inclusion. This heightened competition shifts more value from searchers to builders, as increased bids enhance the value provided to builders. In response, builders strategically adjust their rebate ratios to incentivize searchers to differentiate their bids, thereby improving the builders’ own chances in subsequent block-building auctions. However, our analysis also reveals that this negative feedback—where builders influence searcher bidding through rebate adjustments—is secondary to the direct effect of increased searcher competition. Ultimately, as competition escalates, remaining a searcher becomes less profitable, making the builder role increasingly attractive. These findings underscore that bundle conflicts serve as a key mechanism driving dynamic role-switching behavior within the system.

\section{Discussion}\label{sec:discussion}

The game-theoretic model introduced simplifies specific aspects of the intricate MEV supply chain observed in practice, with its analyses being inherently grounded in and constrained by the model’s assumptions. 
The interactions among profit opportunities available to different MEV participants represent a significant source of externalities. Graph-based approaches provide an intuitive framework for modeling bundle interactions; however, deriving meaningful graph characteristics from real-world data remains challenging. For scenarios where complementarities between submitted bundles are relatively weak, submodularity assumptions can effectively model and capture such externalities \cite{mamageishvili2024searcher}. As \cite{mamageishvili2024searcher} demonstrates, competition among searchers arises when they leverage identical profit opportunities, despite the independence of these opportunities. In our work, we constructed a bundle interaction network and implemented a simplified simulation approach involving a random graph to regulate the conflict probability among bundles. This abstraction transforms complex micro-interactions into the single critical variable of conflict probability, facilitating clearer system-level analyses. However, identifying the characteristics of actual bundle networks poses substantial challenges, which hinder our simulation model’s applicability to empirical data. McLaughlin et al. \cite{mclaughlin2023large} advanced this area by proposing an arbitrage identification algorithm for decentralized exchange applications, where they modeled conflicts between arbitrage opportunities using graphs to determine execution feasibility. Building upon such efforts, we advocate for network science-oriented investigations into the topological structure of bundle networks, informed by real-world blockchain data, as they can yield valuable insights. 

Using game theory to study MEV is crucial for understanding participant behavior and mitigating negative externalities in blockchain systems, as highlighted by the formalization of MEV games and the comparison of transaction ordering mechanisms in related work \cite{mazorra2022price}.
The integration of agents with two strategic approaches and the inherent complexities of the block-building mechanism make it arduous to directly compute the Nash equilibrium for the proposed game-theoretic model. We adopt agent-based simulation methods and genetic algorithms to model the co-evolution of agents, addressing the challenge highlighted in \cite{bartoletti2021towards} regarding agent-based modeling in DeFi systems.
Although we considered various agent learning approaches \cite{brenner2006agent}, we employ genetic algorithms for their strong global search capability via population diversity and stochastic mutation, enabling efficient evolution within a compact strategy set (size 20). Furthermore, our approach incorporates reinforcement learning principles by dynamically updating strategy fitness based on simulation outcomes, rather than relying on a predefined fitness function.

This paper primarily focuses on the strategy selection and value allocation of profit-seeking actors involved in the block-building process. We assume that builders will use the blocks they construct to participate in an equivalent of a second-price auction, without considering any strategic behavior by builders in the block-building auction.
However, a more interesting scenario is that each agent participates as a block builder in the block building auction, while they can also choose to share their bundles to reduce risk. This leads to complex bidding strategies in the block building auction. 
Recently, Wu et al. \cite{wu2024strategic,wu2024compete} 
construct an agent-based model to perform EGTA on the bidding strategies of builders in the block building auction associated with the PBS system.
Combining models of these two distinct stages of the MEV supply chain would induce a more complex game-theoretic problem but could offer deeper insights into understanding PBS systems.

\section{Conclusion}\label{sec:conclusion}

This study introduces a co-evolutionary framework to analyze the strategic behavior of profit-seeking participants in the block-building process within the PBS system. By employing genetic algorithms, we simulate the co-evolution of agents' strategies, enabling us to observe their behavioral dynamics under varying conditions, particularly changes in the probabilities of conflicts between bundles. Through agent-based simulations, we capture the evolutionary outcomes of the system and apply EGTA, specifically the $\alpha$-Rank algorithm, to compute the selection frequencies of agents for two key meta-strategies: block building and bundle sharing.

\bibliographystyle{IEEEtran}
\bibliography{mybibliography}
%


%
\end{document}